\documentclass[useAMS,usenatbib]{mn2e}

\usepackage{epsfig,aas_macros}

\title[A near-IR study of the Sculptor dwarf spheroidal]{A
near-infrared and optical photometric study of the Sculptor dwarf
spheroidal galaxy: implications for the metallicity spread}
\author[C. Babusiaux et al.]
 {C.~Babusiaux$^{1,2}$, G.~Gilmore$^1$, M.~Irwin$^1$\\
$^1$ Institute of Astronomy, University of Cambridge, Cambridge, CB30HA, UK\\
$^2$ Institut d'Astronomie et d'Astrophysique, Universit\'e Libre de Bruxelles, B-1050 Bruxelles}
\begin{document}

\date{Accepted xxx. Received xxx}

\pagerange{\pageref{firstpage}--\pageref{lastpage}} \pubyear{2004}

\maketitle

\label{firstpage}

\begin{abstract}
The Sculptor dwarf spheroidal galaxy has a giant branch with a
significant spread in colour, symptomatic of an intrinsic age/metallicity
spread. We present here a detailed study of the Sculptor giant branch
and horizontal branch morphology, combining new near-infrared photometry 
from the Cambridge
InfraRed Survey Instrument (CIRSI), with optical data from the ESO
Wide Field Imager. 
For a Sculptor-like old and generally metal-poor system, the position of 
Red Giant
Branch (RGB) and Asymptotic Giant Branch (AGB) stars on the
colour-magnitude diagram (CMD) is mainly metallicity dependent. 
The advantage of using optical-near infrared colours is that the position 
of the RGB locus is much
more sensitive to metallicity than with optical colours alone.  In contrast
the horizontal branch (HB) morphology is strongly dependent 
on both metallicity and age. 
Therefore a detailed study of both the RGB in optical-near infread colours 
and the HB can help break the age-metallicity degeneracy.
Our measured photometric width of the Sculptor giant
branch corresponds to a range in metallicity of 0.75 dex.  We detect
the RGB and AGB bumps in both the near-infrared and the optical
luminosity functions, and derive from them a mean metallicity of
[M/H]$ = -1.3 \pm 0.1$. From isochrone fitting we derive a mean
metallicity of [Fe/H]$ = -1.42$ with a dispersion of 0.2 dex.  These
photometric estimators are for the first time consistent with individual 
metallicity
measurements derived from spectroscopic observations.  No spatial gradient is
detected in the RGB morphology within a radius of 13 arcminutes, twice
the core radius.  On the other hand, a significant gradient is
observed in the HB morphology index, confirming
the `second parameter problem' present in this galaxy. These
observations are consistent with an early extended period of star
formation continuing in time for a few Gyr.
\end{abstract}

\begin{keywords}
galaxies: individual: Sculptor dwarf spheroidal -- galaxies: stellar content -- Local Group -- infrared: stars
\end{keywords}

\section{Introduction}

The dwarf galaxies of the Local Group offer a unique opportunity to
quantify the evolution and interactions of low-mass galaxies.  Dwarf
spheroidals were originally thought to be very similar in their
metallicity and star formation histories to the galactic globular
clusters, but their star formation history is now known to be more
complex.  The Sculptor dwarf spheroidal, the first dSph discovered
(Shapley 1938), has a population which is predominantly old and
moderately metal poor (e.g. \citealt{DAC84}, \citealt{MAT98},
\citealt{MON99}, \citealt{DOL02}).  However, the presence of an
extended horizontal branch (e.g. \citealt{DAC84}, \citealt{MAJ99}) and
the detection of associated neutral hydrogen (\citealt{CAR98},
\citealt{BOU03}) suggest the possibility of a more complex star
formation history.  A metallicity spread has been inferred in Sculptor
from the large spread of its red giant branch (e.g. \citealt{DAC84},
\citealt{SCH95}, \citealt{KAL95}, \citealt{MAJ99}) and the period
distribution of RR Lyrae (\citealt{KAL95}, \citealt{KOV01}).
Metallicity gradients have been discovered in several dwarf
spheroidals (\citealt{HAR01}). \cite{MAJ99} and \cite{HUR99} showed
that the red horizontal branch (RHB) of Sculptor is more strongly
concentrated towards the centre than is the bluer population.
\cite{MAJ99} suggest this and the detection of two bumps in the
Sculptor red giant branch are direct evidence for a bimodality in
Sculptor's metallicity distribution. \cite{HUR99} find no radial gradient of the 
age or metallicity distribution within Sculptor, whereas
using the same data, \cite{HAR01} find a metallicity gradient.
Spectroscopic studies of a few stars in the central regions confirmed
a wide abundance range, over 1~dex (\citealt{SHE03}), while first
results from an extensive spectroscopic study of \cite{TOL04} confirm
the broad abundance distribution, with range -2.5$<$[Fe/H]$<$-1.5, and
show that the central regions, where the RHB stars are found, also
contains a significant subpopulation of stars as metal rich as -1~dex.

Photometric studies of the age distribution in Sculptor
(\citealt{MON99}, \citealt{DOL02} and \citealt{RIZ04}) confirm that a
significant metallicity gradient, with some relatively smaller age
spread, are both required by the stellar colour-magnitude diagram, and
are consistent with all other data. This predominately old age is
puzzling, given that Sculptor is apparently unique among the dSph
galaxies in having associated HI. Sculptor is further of interest in
that it may belong to a common orbital plane with several other dSph
(\citealt{LYN76}).

In this paper we present the first near-infrared photometric study of
Sculptor. The combination of our near-infrared data, obtained with the
Cambridge InfraRed Survey Instrument (CIRSI) on the Las Campanas 2.5m
duPont telescope, and optical data from the ESO 2.2m Wide Field Imager
archive, has allowed us to undertake a broader waveband study of the
metallicity spread in Sculptor.  Observations and data reduction
procedures are presented in Section 2.  In Section 3 colour-magnitude
diagrams of Sculptor are compared with theoretical isochrones and
globular cluster data. In section 4 we present the detection of the
RGB and AGB bumps. The metallicity distribution function is derived in
section 5. Section 6 is devoted to the study of the variation of RGB
and HB morphologies with radius.  We conclude with a discussion of the
main results.

\section{The data}

Photometric studies provide a valuable tool to determine internal
metallicity spreads in dSph galaxies. Although individual
spectroscopic measurements are more precise, as yet the numbers of
stars with direct spectra is small. Most such photometric studies use
optical photometry.  The combination of optical and near-infrared data
provides substantially enhanced information, reducing the effect of
photometric errors, allowing colour-colour selection of sources, and
providing in particular the V-K colour, which is a good indicator of
the stellar effective temperature.

We have obtained wide-area near-infrared J and K-band photometry of the Sculptor
dSph galaxy, complementing this with optical V and I-band data from
the ESO archive.

\begin{figure}
\centering
\includegraphics[width=8cm]{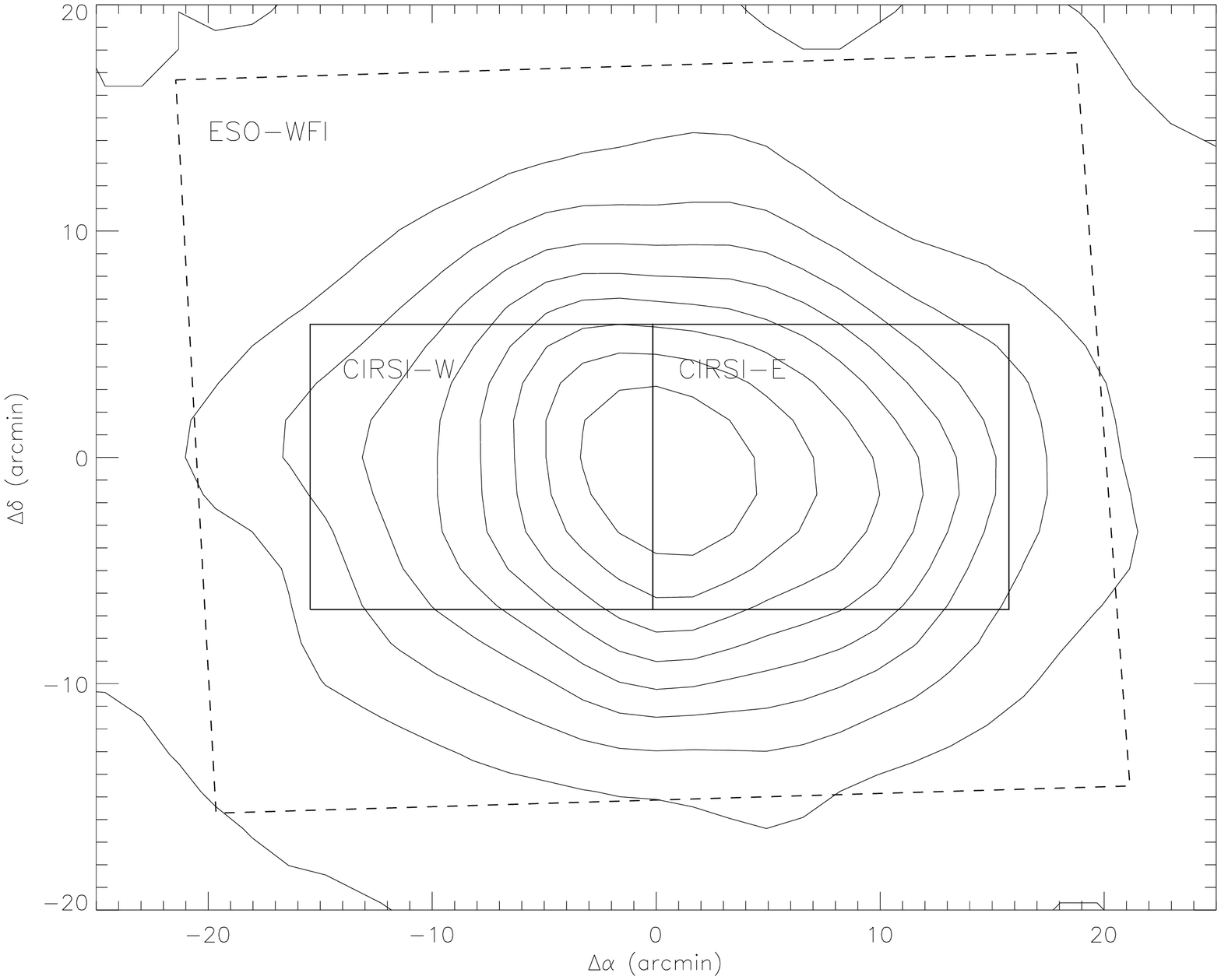}
\caption{Field of view of the CIRSI and ESO-WFI data. The contours are
from Irwin \& Hatzidimitriou (1995). Sculptor's core and tidal radii are 
respectively 5.8 and 75 arcminutes.}   
\label{fov}
\end{figure}

Figure \ref{fov} presents the Sculptor area observed, while 
table \ref{ObsLog} summarises the observations.

\begin{table*}
\caption{Log of the Sculptor observations. The exposure time format (d
x n x exp) lists `n',
the number of short integrations on a specific centre (`loops'), and `d',
the number of slightly offset (`dither') pointings.
The seeing conditions, measured from the images, are indicated by the PSF FWHM (in
arcsec) with the corresponding range during the mosaic observations with
CIRSI.} 
\begin{tabular}{ll|lllll}
\hline
Instrument & Field & Filter & Date & Exposure (secs) & seeing (\arcsec) \\
\hline
\hline
CIRSI & Sculptor-West & J & 2001-09-04 & 9 x 5 x 20 & 1.02-1.4 \\
 & & K$_s$ & 2001-09-03 & 9 x 3 x 45 & 0.8-0.9 \\
\hline
  & Sculptor-East & J & 2001-09-04 & 9 x 5 x 20 & 1.14-1.4 \\
 & & K$_s$ & 2001-09-03 & 9 x 5 x 20 & 0.96-1.3 \\
\hline
\hline
ESO WFI & Sculptor & V & 1999-07-22 & 3 x 300, dit-3 & 1.00 \\
 & & I & 1999-07-22 & 3 x 300 & 0.91 \\
\hline
\end{tabular}
\label{ObsLog}
\end{table*}

\subsection{The near-infrared data}

Near-infrared observations in the $J$(1.25$\mu$m) and
$K_s$(2.15$\mu$m) bands were made with the Cambridge Infrared Survey
Instrument (\citealt{BEC97}, \citealt{MAC00}) on the du Pont 2.5m at
Las Campanas Observatory.  CIRSI is a mosaic imager consisting of four
Rockwell 1k x 1k detectors. The pixel scale is 0.2 arcsec
pixel$^{-1}$.  The gaps between the detectors being comparable to the
detector size, four dither sets are needed to fill the mosaic image,
leading to a field of view of about 13 x 13 arcmin$^2$.  For each
dither set, 9 dither frames are taken with offsets of about 10
arcsec. Each dither frame is observed in 5 loops of 20 seconds
exposure time each. The total exposure time per mosaic is then 900s.
Two mosaics were taken at the centre of the Sculptor dwarf spheroidal,
as pictured in figure \ref{fov}. 

\subsubsection{CIRSI data reduction} 

The data reduction was made using an updated version of the InfraRed
Data Reduction (IRDR) software package, first developed by
\cite{SAB01}. A summary of the full process is given here. The updated
version of IRDR with its documentation are available at
http://www.ast.cam.ac.uk/\~{}optics/cirsi/software.  A fuller
description is provided by \cite{BAB05}, in their presentation of a
CIRSI study of the Galactic bulge and bar.

First each image is corrected for non-linearity.  The dark current for
the relevant exposure time is then subtracted.  The data are flatfield
corrected using lamp-on domeflats subtracted with lamp-off domeflats
and normalized to the first detector sensitivity.  These flatfields
are also used to detect bad pixels and create weight maps, used during
the dither frame coaddition.

The sky is subtracted in two passes. A first pass sky image is made by
median combining the nearest loop-combined frames of the dither set.
After a first dither frame coaddition, object masks are produced using
SExtractor source extraction (\citealt{BER96}). A masked frame is
created from this source-detection list, with an enlarged area around
each detected source being used. This object-masked frame is used to
make a second pass sky subtraction on each loop.

Spatial offsets between loop-combined dither frames are computed by
cross-correlating object pixels mapped by SExtractor.  Dither frames
are then coadded using a weighted bi-linear interpolation, excluding
bad pixels.

Finally, the astrometry is calibrated by correlating the SExtractor's 
object catalogue with the USNO-A2 catalogue.

\subsubsection{CIRSI photometry}

Once the data are reduced, we use the IRAF photometry routines.
Source are detected using the IRAF {\sc daofind} procedure, with a
significance threshold set at five-sigma.  Aperture photometry is then
obtained using the IRAF {\sc phot} task.  The aperture radius is
assigned for each dither frame to the measured PSF FWHM.  
Observations of standard stars from \cite{PER98} were
obtained each night, and were used to derive the magnitude zero-point
of each night. Standard star photometry used an aperture photometry
radius of 20 pixels, equivalent to a diameter of 8 arcsec. The
internal zero point dispersion derived from multiple observations of
the standard stars during a night is 0.013 mag in J and 0.008 mag in
K.  The instrumental magnitudes derived using the psf-fwhm aperture
are corrected for aperture effects using the curve-of-growth method
(\citealt{STE90}), implemented with the IRAF {\sc mkapfile} task
applied to selected bright isolated stars.  Airmass corrections of
$k_J$=0.1 and $k_K$=0.08 mag per air-mass (\citealt{PER98}) are
applied to the photometry.  Finally, images detected near the borders
of the images are eliminated, so that only stars observed in all the
dither frames are kept. False detections located in the wings of
highly saturated stars are manually deleted.

The photometric calibration was checked by correlating the brightest
stars with the 2MASS catalogue. The CIRSI K$_s$ photometry is
consistent with the 2MASS photometric system.  However, an offset in
the J-band photometry of $J_{(CIRSI)}-J_{(2MASS)} = -0.042 \pm 0.005$
is observed.  This is due to the 2MASS J-band filter being more
extended into the atmospheric water absorption features at around 1.1
and 1.4 $\mu$m (\citealt{CAR01}) than is the filter used for CIRSI.

The 5-$\sigma$ magnitude completeness limits are about J$\sim$20 and
K$_s$$\sim$18.8. The photometric errors are about 0.025 mag for
J$<$17, 0.08 mag at J=20, 0.022 mag for K$_s<$16, and 0.08 mag at
K$_s$=18.8.

\subsection{The optical data}

The optical data were obtained from the ESO 2.2m telescope Wide Field
Imager archive. They consist of 3x300s dithered exposures in V and I.
Each of the eight 4kx2k CCDs in the ESO 2.2m WFI covers around 8
arcmin x 16 arcmin of sky at a sampling of 0.238 arcsec per pixel,
comparable with the CIRSI data.  The total field of view of the WFI is
about 33 arcmin x 33 arcmin, including small gaps of around 10 arcsec
between CCDs. This field of view entirely covers the observed CIRSI
fields (figure \ref{fov}).

The WFI data were processed using a variant of the standard optical
pipeline described by \cite{IRW01}.  After trimming, bias-correcting,
flatfielding and gain normalisation, the I-band exposures were
additionally defringed using a I-band fringe frame derived from the 3
I-band science images. To generate the fringe frame, the 3 dithered
I-band exposures were object masked, combined with rejection to remove
residual artifacts, and then further filtered to improve the local
signal-to-noise ratio.

Object catalogues were generated for each individual processed science
frame and used to define, and refine, the World Coordinate System
(WCS) for each frame, by comparison with the online APM plate
catalogues.  After updating the 2D images with the derived WCS, the 3
V and 3 I frames of Sculptor were stacked with cosmic ray rejection,
using the WCS for coalignment, and confidence maps, derived from the
flatfield frames and bad column lists for each CCD, to aid in
rejection.

Final object catalogues were then derived from the stacked frames, and
used both to update the WCS and to provide morphological
classification information for each detected object.  Object detection
proceeds via a search for contiguous pixels at a threshold above the
local sky; following detection a series of object parameters are
derived.  These latter are used to generate position, flux and shape
information for use in later processing stages (\citealt{IRW85},
\citealt{IRW97}).  The basic photometric measurement used is an
aperture flux estimate with radius set to the average FWHM in the
stacked frames. Additionally, all detected images have a series of
fixed aperture measures produced (scaled to the basic aperture size
used) in order to automatically derive aperture corrections for
stellar images for each CCD.

Archive observations of \cite{LAN92} standard fields define native
system zero-points in each passband using the colour equations for WFI
available on the ESO web site. The gain-correction in the previous
stage ensures that all CCDs are on the same internal system,
normalised to CCD1.  However scattered light leads to a variation of
the photometric zero points across the mosaic (\citealt{MAN01}).  From
the standard field observations, a correction term of 1.5r$^2$, with r
being the distance in degrees from the optical axis, was found to be a
good approximation of the effect of the scattered light. Overall, this
provides a zero-point calibration with 1-2\% accuracy.

The 5-$\sigma$ magnitude completeness limits are about V$\sim$24,
I$\sim$22.5, so that the censorship on the data is due only to the
CIRSI J and K$_s$ photometric limits.  The errors are smaller than
0.01 mag in V and 0.02 in I for magnitudes brighter than V$\sim$21.5
and I$\sim$19.5. 

\section{Photometry of the Sculptor dSph galaxy}

\begin{figure}
\centering
\includegraphics[width=8cm]{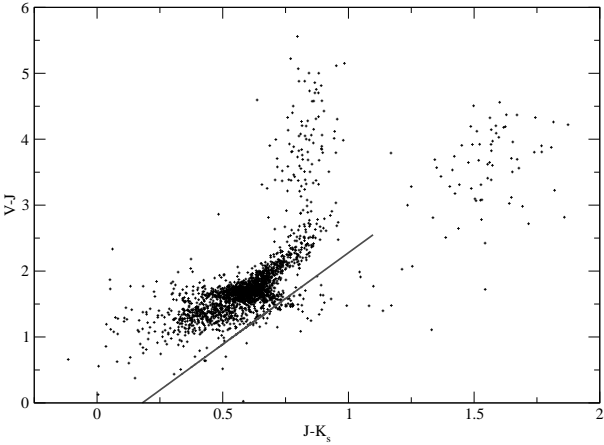}
\caption{The V-J vs J-K$_s$ colour-colour diagram of all point sources
detected in V,I,J and K$_s$. The solid line shows the quasar selection
criterion from the KX method (Warren et al. 2000).}
\label{colourtoclean}
\end{figure}

Our next task is to generate a list of point-sources which are stellar
members of the Sculptor dSph galaxy. Extended objects are readily
eliminated from further consideration using the morphological flags
from the optical pipeline.

\subsection{Selection of Sculptor stars}
The VJK colour-colour diagram, figure \ref{colourtoclean}, allows the
detection of three other point-source populations unrelated to the
Sculptor dSph galaxy.  Red quasars are clearly seen with colours which
are too red to correspond to any star. All sources with J-K$_s>$1 were
eliminated.  A stream of stars with J-K$_s$ about 0.8~mag and V-J
colour redder than 3~mag can be seen. They are foreground low-mass
dwarfs (e.g. \citealt{LEG92}); these can be eliminated, without
discarding any Sculptor stars, by excluding all stars redder than the
colour of the tip of the Sculptor giant branch in all relevant
colour-colour CMDs (e.g. V-I$>$1.9).  A number of probable blue
quasars can be seen bluer in V-J than the main stellar locus of Sculptor and
foreground stars. These are selected and eliminated according to the
KX method criterion (\citealt{WAR00}), which is shown as the solid
line in figure \ref{colourtoclean}.  Remaining foreground Galactic
stars are minimised by our selection inside each colour-magnitude
diagram of the location of Sculptor member stars, as described further
below.

\begin{figure}
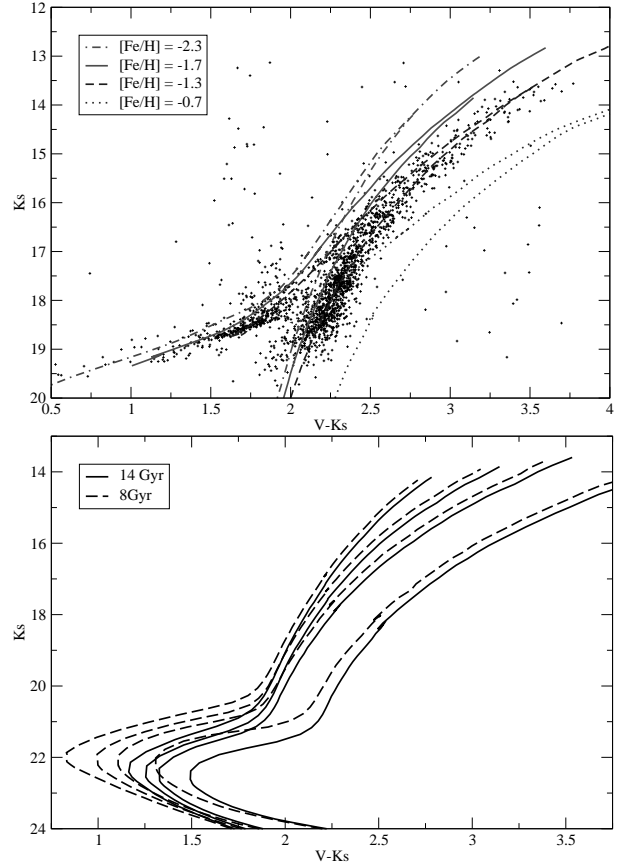

\centering
\includegraphics[width=8cm]{fig3a.eps}
\hspace{0.1\textwidth}
\includegraphics[width=8cm]{fig3b.eps}
\caption{Theoretical isochrones of the Padova group
(a) AGB and RGB isochrones for an age of 14 Gyr and for metallicities
  [Fe/H]=-2.3,-1.7,-1.3,-0.7 dex from left to right. The data points are
our photometry. 
(b) The RGB and main sequence turn-off for ages 14 and 8 Gyr and for
  the same four metallicities as in the top panel.}    
\label{isochrones}
\end{figure}

\subsection{The Sculptor Colour-Magnitude diagrams}
It is apparent from figure \ref{isochrones}a. that there is a
significant real width to the RGB of Sculptor in the (V-K$_s$,K$_s$)
colour-magnitude diagram, confirming several previous studies of Sculptor CMD. 
The RGB photometric width is about
$\Delta(V-K_s)=0.3$ at magnitude K$_s$=16, where the photometric
measurement errors are 0.023 mag.

To determine the origin of this dispersion, theoretical isochrones
from the Padova group, given in the ESO-WFI and 2MASS photometric
systems by respectively \cite{GIR02} and \cite{BON04}, have been
overlaid on the Sculptor's CMDs, using a distance modulus
(m$-$M)$_0$=19.54 and an extinction E(B-V)=0.02 (\citealt{MAT98}). The
extinction is derived in the different photometric bands using the
\cite{CAR89} extinction curve.

Figure \ref{isochrones}b shows theoretical RGB isochrones for
metallicities Z=0.0001, 0.0004, 0.001 and 0.004 with solar mixture,
and ages 8 and 14 Gyr.  It can be seen that the RGB is much more
sensitive to metallicity than to age. The main sequence turn-off,
which is more sensitive to age, was used by \cite{MON99} to derive an
age of 15$\pm$2 Gyr for Sculptor. The results of \cite{DOL02} and
\cite{RIZ04} confirm the predominance of old stars in Sculptor.  We
therefore adopted isochrones of age 14 Gyr for figure
\ref{isochrones}a. These isochrones clearly confirm the presence of a
metallicity spread within the Sculptor RGB stars.

\begin{figure}
\centering
\includegraphics[width=8cm]{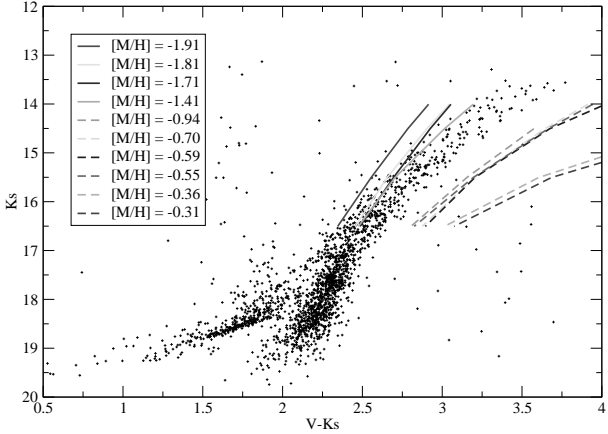}
\caption{Sculptor photometric data from this study (points), together
with the fiducial RGB lines for the sample of Galactic globular
clusters from Ferraro et al. (2000), with their global metallicities as
defined by Ferraro et al. (1999) indicated.} 
\label{ferraro}
\end{figure}

The colour magnitude diagrams can also be compared directly to
globular cluster observations.  \cite{FER00} and \cite{SAV00} provide
fiducial lines for the RGB of Galactic globular clusters for a wide
range of metallicity, in the (V,J,K) and (V,I) photometric systems.
Figure \ref{ferraro} shows the fiducials of \cite{FER00} globular
clusters on the (V-K$_s$,K$_s$) colour-magnitude diagram of
Sculptor. The transformation from absolute to relative magnitudes is
the same as the one applied for the theoretical isochrones.  However
the V and K magnitudes are not on the same photometric system as ours,
leading to expected photometric differences of the order of 0.1 mag.
Considering that globular clusters tend to have alpha-enhanced 
element ratios, whereas the Sculptor stars do not, we indicate their global
metallicities as defined by Ferraro et al. (1999). The definition of this 
global metallicity scale and how it can be translated into [Fe/H] for Sculptor
is discussed in section 4.
Here again a metallicity spread is confirmed as being consistent with
the width of the Sculptor RGB. 
Figure \ref{ferraro} illustrates that all stars in Sculptor are more 
metal-poor than [M/H]$=-1.0$. 

The spread of the RGB in the (J-K$_s$,K$_s$) and (V-I,I) CMDs is
smaller and more sensitive to photometric errors than in
(V-K$_s$,K$_s$). According to \cite{SAV00}, a variation of metallicity
from $-2.0$ to $-1.5$ dex results in a difference in V-I of 0.04 mag
at I=$-2$ (one magnitude brighter than the RGB bump), while according
to \cite{FER00}, the same variation of metallicity at K=$-3$ results
in a variation of 0.2 mag in V-K. Considering the relative photometric
errors in V, I and K$_s$, this means that V-K$_s$ is 1.7
times as sensitive to metallicity as is V-I. We will therefore use 
preferentially the (V-K$_s$,K$_s$) CMD in the following to derive 
photometric metallicity indicators. 

\section{The RGB and AGB bumps}

\begin{figure}
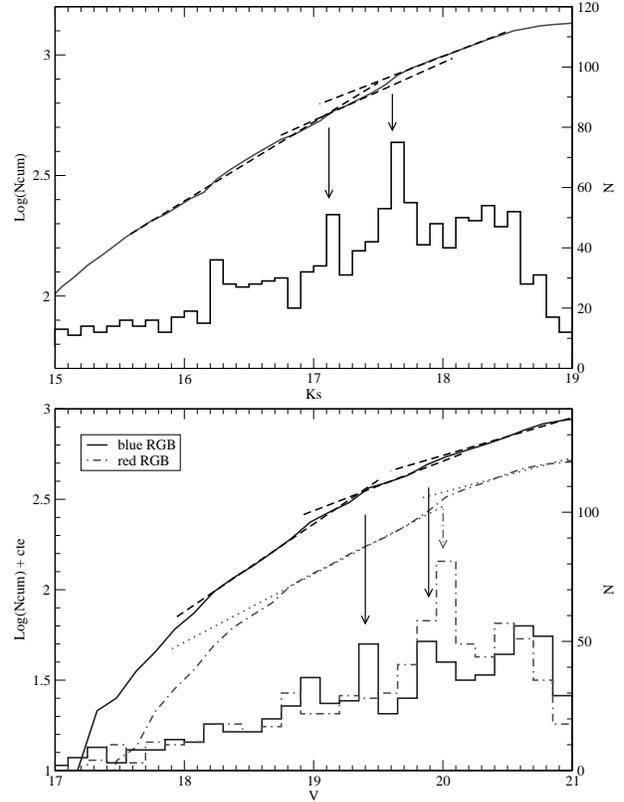

\centering
\includegraphics[width=8cm]{fig5a.eps}
\hspace{0.1\textwidth}
\includegraphics[width=8cm]{fig5b.eps}
\caption{Luminosity Function (LF) and cumulative LF (a) in the K$_s$
band for all stars, (b) in the V band separately for stars redder and
bluer than the giant branch mean fiducial.}
\label{bump}
\end{figure}

Local maxima are observed in the luminosity function of the giant
branch of old metal-poor stellar populations. These RGB and AGB bumps
are well known features of the colour-magnitude diagrams of globular
clusters (e.g. \citealt{FER99}). Those two bumps are detected in all our 
visible and near-infrared bands (table \ref{bumptab}), as illustrated in 
figure \ref{bump} for the V and K$_s$ bands. To allow us to compare the
absolute magnitudes of these features in Sculptor with other studies,
the ESO WFI V and I photometry was converted into the standard Johnson
photometry using the \cite{GIR02} isochrones in those two filter systems. 
At the location of the bumps, $(V-I)_{J}=0.96$, a given simulated star of 
the isochrones present the colours $V_J-V_{WFI}=-0.06$ and
$I_J-I_{WFI}=0.1$. As previously, the conversion to absolute
magnitudes assumes a distance modulus of 19.54 mag and
E(B-V)=0.02. Errors in the determination of the bump location are of
about 0.1 mag.
  
\begin{table*}
\centering
\caption{Magnitudes of the RGB and AGB bumps. See the text for a
definition of the photometric system.}
\begin{tabular}{l|cccc|cccc}
\hline
  & $m_{V_{WFI}}$ & $m_{I_{WFI}} $ & $m_J$ & $m_{K_s}$ 
  & $M_{V_J}$ & $M_{I_J}$ & $M_J$ & $M_{K_s}$ \\
RGB bump & 19.92 & 18.83 & 18.23 & 17.61 & 0.26 & -0.64 & -1.33 & -1.94  
\\
AGB bump & 19.40 & 18.38 & 17.75 & 17.13 & -0.26 & -1.09 & -1.81 & -2.42 
 \\
\hline
\end{tabular}
\label{bumptab} 
\end{table*}

\cite{MAJ99} also detected those two bumps within the central 10\arcmin, 
but associated the second
one with the RGB bump of a more metal-poor population. Indeed, the AGB
bump of a population of metallicity [Fe/H]$\simeq -1.5$ is located at
the same position on the CMD as is the RGB bump of a population of
metallicity [Fe/H]$\simeq -2$. The V magnitude of the second bump is
consistent with the value of $M_V(AGBbump) = -0.3 \pm 0.1$ used by
\cite{FER99}.  Its clear detection can be explained by the fact that
the luminosity level of the AGB bump stays fairly constant with the
cluster metallicity (e.g. \citealt{CAS91}).  The AGB bump being always
bluer than the RGB, it explains the detection of this second bump on
the blue side of the RGB by \cite{MAJ99}. Using, as they did, a division of the
giant branch into a red and blue part (figure \ref{fiduc}),
we not only find as expected the second bump on the blue side of the
giant branch, but also the RGB bump shifted by about 0.2 magnitudes
(figure \ref{bump}b), which is consistent with a variation in
metallicity. If the second bump is to be due to a distinct metal-poor
population, being clearly detected at all wavelengths, this second
population should also show a clear imprint on the tip of the red
giant branch. No such distinct second RGB can be detected (figures
\ref{isochrones}a. and \ref{ferraro}).  We then conclude that the
second bump is the AGB bump.

The V magnitudes of the HB and the RGB bump are good indicators of the
 metallicity (e.g. \citealt{FER99}).  The peculiar shape of Sculptor's
 HB however makes the determination of its mean V magnitude unreliable
 (figure \ref{hbindex}).  
The V magnitude of the RGB bump leads to a global metallicity of
[M/H]$=-1.30\pm$0.12 from the calibration of \cite{FER99}. Its
K$_s$ magnitude leads to [M/H]$=-1.39\pm0.14$ according to
\cite{CHO02}, while it leads to [M/H]$=-1.19\pm0.12$ from the new
calibration of \cite{VAL04}.  
Those calibrations were made using the relation of \cite{SAL93}: 
\begin{equation}
[M/H] = [Fe/H] + \log(0.638*10^{[\alpha/Fe]}+0.362)
\label{eq:sal93}
\end{equation} 
Sculptor does not seem to present a strong enhancement of alpha-elements:
\cite{SHE03} measured for 5 stars the following values for [$\alpha$/Fe]: 
0.18, 0.13,0.13,-0.01 and 0.23 (table 2 of \citealt{TOL03}). 
For [$\alpha$/Fe]=0.13, [M/H]=-1.3 can be translated to [Fe/H]=-1.4
 by equation \ref{eq:sal93}.

It should be stressed that the indicated errors do not take into account the 
uncertainty in the distance modulus, estimated to be 0.08 mag in \cite{MAT98}.
Another metallicity indicator, independent of the distance modulus and
of zero point calibration errors, is the difference between the AGB
and RGB bump luminosities. From \cite{FER99}, with
${\delta}V^{RGBbump}_{AGBbump}=0.52\pm0.14$, we can derive
[M/H]=$-1.4\pm0.2$.  

Those metallicity indicators agree with the
previous comparisons of the RGB morphology with theoretical isochrones
and globular clusters (figure \ref{isochrones}a and figure
\ref{ferraro}).

\section{The metallicity distribution function}

\begin{figure}
\centering
\includegraphics[width=8cm]{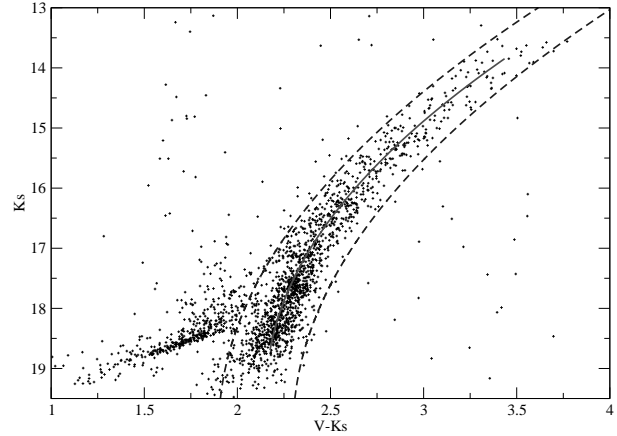}
\caption{Least square fit of a second order polynomial to the Sculptor
giant branch. The dashed curves frames the stars selected for the
giant branch studies.}
\label{fiduc}
\end{figure}

\begin{figure}
\centering
\includegraphics[width=8cm]{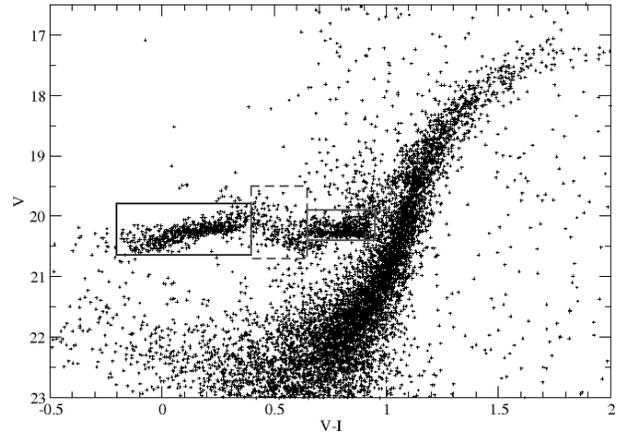}
\caption{Sculptor (V-I,V) CMD. The boxes indicate how the HB index was
computed, selecting the B,V and R stars.} 
\label{hbindex}
\end{figure}

The mean fiducial of the Sculptor red giant branch was computed
through a least squares fit to a second order polynomial on the
(V-K$_s$,K$_s$) CMD.  Horizontal branch stars and foreground stars
have been eliminated by selecting only the stars 0.2 magnitudes away
from the mean fiducial (figure \ref{fiduc}).  Only the giant branch
stars brighter than K$_s$=18.7 and up to the RGB tip (K$_s$$>$13.8)
have been selected.  As the AGB stars occupy the same location on the
CMD as the most metal poor RGB population, no attempt to discard AGB
stars was made.

We derive the metallicity distribution of those selected Sculptor
giant branch stars using the Padova isochrones described
previously. Each isochrone of figure \ref{isochrones}a can be
approximated by a second order polynomial:
\begin{equation}
V-K_s = a_0 + a_1 K_s + a_2 {K_s}^2
\label{eq:fidiso}
\end{equation}
A second order polynomial regression of those coefficients as a
function of the metallicity of the isochrones is obtained:
\begin{equation}
a_i = a_{(i,0)} + a_{(i,1)} [Fe/H] + a_{(i,2)} {[Fe/H]}^2
\label{eq:fidmet}
\end{equation}
By inverting equation \ref{eq:fidiso}, each point in the
(V-K$_s$,K$_s$) CMD can then be assigned an estimate of its
metallicity. Taking into account only the uncertainty of the accuracy 
on the polynomical regression of equations \ref{eq:fidiso} and \ref{eq:fidmet}
and the photometric errors, the typical uncertainty in the resulting 
measurement of the metallicity of a individual star is smaller than 0.04 dex. 

\begin{figure}
\centering
\includegraphics[width=8cm]{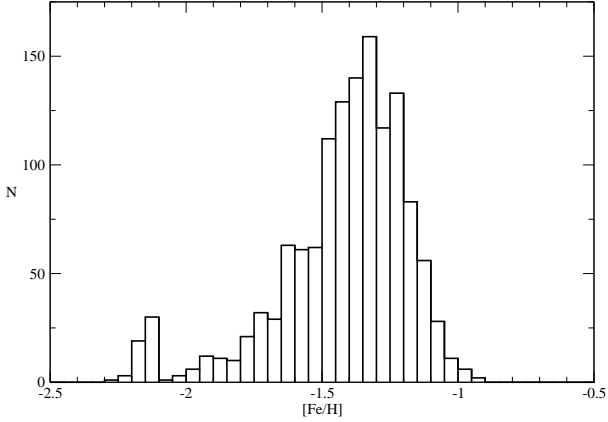}
\caption{Photometric metallicity distribution function as derived
using theoretical isochrones.}
\label{fig:mdf}
\end{figure}

The resulting metallicity distribution function is represented in
figure \ref{fig:mdf}.  The secondary peak at [Fe/H]$= -2.2$ is an
artefact due to AGB stars and should be ignored.  The mean metallicity
obtained is [Fe/H]$ = -1.42$ with a dispersion of 0.2 dex.  This mean
metallicity is in agreement with the value obtained from the RGB
bump. A comparison of those metallicity 
estimates with the literature is given in the discussion section.

\section{Metallicity gradient indicators}

\subsection{The Giant Branch}

Since the RGB in our photometric system provides a good indicator of
metallicity, we studied its variation with radius as a test of a
possible metallicity gradient in Sculptor.

\begin{figure}
\centering
\includegraphics[width=8cm]{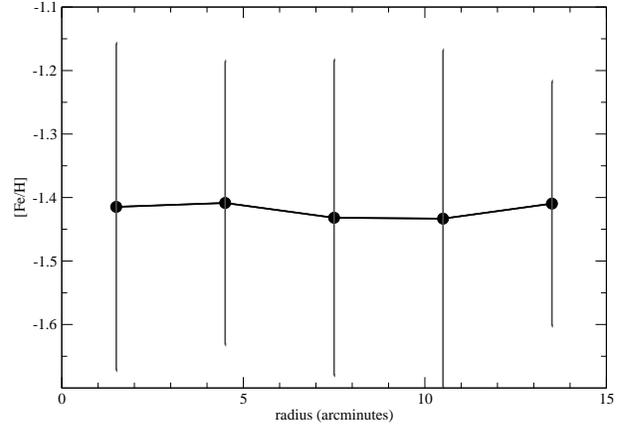}
\caption{Mean metallicity, as estimated in section 5, as a function of radius. 
The vertical line show the measured dispersion around this value.}
\label{fig:metrad}
\end{figure}

The metallicity indicators derived in the previous section have been
studied as a function of radius. The central 10 arcminutes of Sculptor
has zero ellipticity (\citealt{IRW95}), so as our data do not extend
beyond a radius of 15 arcminutes, we use a simple circular annulus.
Figure \ref{fig:metrad} show that no metallicity gradient is detected 
within our data. This gives an upper limit of 0.03 dex for the metallicity 
gradient within twice the core radius of Sculptor. 

\subsection{The Horizontal Branch}

The horizontal branch of Sculptor can be clearly divided into a blue (B)
and a red (R) part, lying on either side of the instability strip (V). 
The ratio of the number counts of those different parts, quantified by
the HB index (B-R)/(B+V+R), is dependent on the metallicity. But there
is a well known `second parameter problem', which could be age
(e.g. \citealt{LEE94}).   
 
\begin{figure}
\centering
\includegraphics[width=8cm]{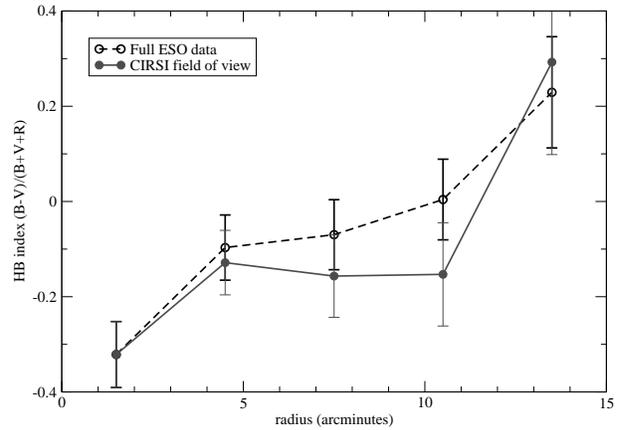}
\caption{The HB index against radius for all the ESO-WFI data and
  stars only within the CIRSI field of view (figure \ref{fov}).} 
\label{hbrad}
\end{figure}

The HB index was computed from the (V-I,V) CMD, as illustrated in
figure \ref{hbindex}, for the same radial annuli as used for the RGB
study. Both the full ESO WFI field of view data and the sub-area in
common with the CIRSI field of view are presented in figure
\ref{hbrad}.  A K-S test for the hypothesis that the red and blue
horizontal branch stars have the same radial distribution gives a
significance level of 10$^{-9}$\%.  This confirms the HB gradient
detected by \cite{HUR99} and \cite{MAJ99}.

The \cite{LEE94} models give theoretical isochrones that show, for an HB
index between -0.5 and 0.5 and a given age, a linear
relation between the HB index and the metallicity:
\begin{equation}
[Fe/H] \simeq -0.34 * HBindex + cte
\label{hbmet}
\end{equation}
The observed gradient in HBindex of about 0.5 then corresponds to a
gradient in metallicity of 0.17 dex. Considering the upper limit of 0.04
dex for a metallicity gradient derived previously from the RGB morphology,
an age gradient is required to explain the observed HB gradient.
As always when discussing HB
morphology however, one must recall that the `second parameter
problem' is not yet solved and that another parameter may influence
the HB morphology.

The simplest conclusion is that a small gradient in mean age is
apparent in Sculptor and that an eventual small metallicity gradient
associated with it would have a too small effect on the RGB compared
to the large abundance dispersion to be detected.

\section{Conclusions and discussion}

The combination of near-infrared photometry from CIRSI with optical
data from the ESO WFI, allowed a detailed study of the Sculptor dwarf
spheroidal giant branch morphology.  We confirm that the broad giant branch of
Sculptor demonstrates an intrinsic metallicity spread 
(e.g. \citealt{DAC84}, \citealt{SCH95}, \citealt{KAL95}, \citealt{MAJ99}). 
From our photometric study we quantify this spread into a metallicity range of
$\Delta[Fe/H]=0.75$ dex.  The RGB and AGB bumps are detected in all
the optical and near-infrared luminosity functions, excluding the
substantial metal-poor contribution to Sculptor's metallicity
distribution proposed by \cite{MAJ99}. We derive a mean metallicity
within two core radii in Sculptor of about [Fe/H]$=-1.4$ from both the RGB and AGB 
bumps magnitudes and isochrones fitting.
Our mean metallicity and the metallicity range are higher than those derived in
\cite{DAC84} from photometry of the RGB and in \cite{KAL95} from RR
Lyrae stars: these results were summarized in the \cite{MAT98} review
as a mean of [Fe/H]$=-1.8\pm$0.1 with a spread of 0.3 dex.  However our
photometry is in agreement with the metallicity estimations from
Sculptor RR Lyrae stars of \cite{KOV01} and CaII triplet observations
of 37 stars of \cite{TOL01}. It is in excellent agreement with the
very recent spectroscopic survey of \cite{TOL04}, whose derived
metallicity distribution inside two core radii 
indicates a mean metallicity of $-1.4$~dex, and metallicity range of
about 1~dex. Their data show the population structure is complex, with
the more metal-poor part of the distribution function becoming
dominant at radii beyond those we have studied here.

We do not detect a gradient in the RGB morphology within a radius of
13\arcmin, 2.2 times the Sculptor core radius. Although \cite{HAR01} and 
\cite{TOL04} find a metallicity gradient in Sculptor, our result is in 
agreement with their data.  Indeed figure 6 of \cite{HAR01} shows that
the radial distribution of blue and red RGB stars begins to differ only after
13\arcmin. Figure 3 of \cite{TOL04}, based on spectroscopic data, 
indicates that a metallicity gradient is indeed visible only beyond this 
radius. 
On the other hand, we do detect at high significance a gradient in the horizontal branch
(HB) morphology, confirming the results of \cite{HUR99} and \cite{MAJ99}. 
As this cannot be explained by metallicity, the most
likely second parameter could be age.  \cite{HUR99} did not find
evidence for a gradient in the main-sequence turn off, leading to an
upper limit of a 2 Gyr variation at constant metallicity. According to
the models of \cite{LEE94} a small variation in age of even a few Gyr
leads to a strong variation in the HB morphology. Age could then still
be the second parameter. Moreover, [Fe/H] and age are not the only
variables which can affect the HB morphology, but also other element
abundances, in particular the [O/Fe] ratio (\citealt{LEE94}).
\cite{HUR99} detected another population presenting a gradient in
Sculptor: a `spur' of stars extending $\sim$0.7 mag above the old main
sequence turn off. They conclude that it cannot be explained by the
presence of an intermediate age population, preferring the
interpretation of the spur as a binary sequence, and speculate that it
could be related to the variation of the HB morphology.

No significant intermediate age population has been found in Sculptor,
excluding star formation within the last 5~Gyr.  The large metallicity
spread requires extended star formation, while the evidence of low
alpha-element enhancement implies extended star formation and
self-enrichment over a period of at least 1~Gyr: quite long enough to
affect the HB morphology.  The HB morphology gradient implies that the
most recent star formation episode occurred in the centre of the
galaxy, consistent with naive expectation that gas is more easily
retained deeper in the galaxy potential well. 
Indeed in most dwarf galaxies observed with sufficiently deep wide-field imaging, 
populations gradients have been found with the younger stars being more centrally
concentrated (e.g. \citealt{SAV01} and references therein).
Our lack of detection of a metallicity gradient may be explained by the age-metallicity
degeneracy that would hide a small age and metallicity difference and
by the stronger dependence of the HB on other parameters such as the
oxygen abundance, stellar rotation or binarism, 
as already suggested by \cite{HUR99} and \cite{MAJ99}.

All this could be consistent for Sculptor with a single period of star
formation extended in time for of the order of a few Gyr.
\cite{TOL03} conclude that their study of the element abundances are
consistent with a closed-box chemical evolution scenario.  The small
dynamical mass of dwarf spheroidals such as Sculptor means that their
binding energy is small compared to the energy released by several
supernovae, which leads the high metallicity spread and relatively
high mean metallicity derived for Sculptor puzzling: how did the gas
stay bound long enough to have an extended star formation and gas
enrichment? The star formation rate should be low to allow the
chemical enrichment to proceed gradually. Hydrodynamical simulations
are trying to answer this question (e.g. \citealt{CARR01},
\citealt{CAR02}).

\section*{Acknowledgments}
We are grateful to Jacco van Loon for his assistance with the CIRSI observations.  
We thank the anonymous referee for helpful comments.
The development and construction of CIRSI was made
possible by a grant from the Raymond and Beverly Sackler Foundation.

\bibliographystyle{mn2e}

\label{lastpage}

\end{document}